\documentclass[12pt]{iopart}

\usepackage{graphicx}
\usepackage{amssymb}
\usepackage{float}
\usepackage{xcolor}

\begin{document}
\title[Simultaneous drop in mean free path and carrier density in high-$T_{\rm c}$ cuprates]{Simultaneous drop in mean free path and carrier density at the pseudogap onset in high-$T_{\rm c}$ cuprates}

\author{J G Storey$^{1,2}$}

\address{$^1$ Robinson Research Institute, Victoria University of Wellington, P.O. Box 600, Wellington, New Zealand}
\address{$^2$ MacDiarmid Institute, Victoria University of Wellington, P.O. Box 600, Wellington, New Zealand}
\ead{james.storey@vuw.ac.nz}

\begin{abstract}
High-temperature superconducting cuprates are distinguished by an enigmatic pseudogap which opens near optimal doping where the superconducting transition temperature is highest. Key questions concern its origin and whether it is essential in any way to superconductivity.
Recent field-induced normal-state transport experiments on hole-doped cuprates have measured abrupt changes in the doping dependent Hall number and resistivity, consistent with a drop in carrier density from $1+p$ to $p$ holes per copper atom, on entering the pseudogap phase. In this work the change in resistivity is analyzed in terms of an antiferromagnetic-order-induced Fermi surface reconstruction model that has already successfully described the Hall number. In order for this model to describe the resistivity we find that the zero-temperature mean free path must also drop abruptly in proportion to the size of the Fermi surface. This suggests that intrapocket scattering underlies the observed upturn in resistivity in the pseudogap state.
\end{abstract}
\vspace{2pc}

\submitto{\SUST}
\maketitle

\section{Introduction}
One of the most puzzling details of hole-doped cuprate high-temperature superconductors is the origin of a partial gap in the electronic spectrum known as the pseudogap that coexists and competes with superconductivity. It has long been argued that the pseudogap opens at a critical doping $p^*$ (also denoted $p_{crit}$ and $p_c$) within the superconducting phase\cite{OURWORK1}. This is supported by abrupt drops in the doping dependence of several ground-state properties. These include the condensation energy\cite{ENTROPYDATA2}, zero-temperature superfluid density\cite{ENTROPYDATA2,BERNHARD,STOREYRAMAN,ANUKOOLPRB}, the critical zinc concentration required for suppressing superconductivity\cite{TALLON4,STOREYRAMAN} and the zero-temperature self-field critical current\cite{NAAMNEH}. It has now become clear that on crossing $p^*$, the Fermi surface undergoes a change in topology  from a large hole-like barrel centered at ($k_x$,$k_y$) = ($\pi$,$\pi$) to small arcs terminating at the zone diagonal along (0,$\pi$) to ($\pi$,0)\cite{Fujita09052014}. The so-called Fermi arcs can be interpreted as the visible portion of a closed hole-like pocket arising from reconstruction of the large Fermi surface, for which there are several proposed models in the literature, for example \cite{CHUBUKOV,CHAKRAVARTY,YRZ,HARRISON,SACHDEV,AMPEREAN}. Although there is some evidence from angle-resolved photoemission spectroscopy for curvature at the tips of the arcs expected from pockets\cite{YANG2}, the arcs have also been interpreted as thermally broadened nodes\cite{FERMIARCS2,REBER2012}. Nevertheless, various physical properties in the pseudogap phase have been shown to be consistent with the collapse in size of the Fermi surface\cite{STOREYRAMAN,STOREYENTROPY,LEBLANC1,LEBLANC2,CARBOTTE,RICE}.

The ability to perform transport and thermodynamic experiments at increasingly high magnetic fields is providing unprecedented access to the normal-state at low temperatures by suppressing superconductivity. High-field Hall effect measurements reveal that the Hall number, a measure of the carrier density, transitions from $1+p$ to $p$ holes/Cu as the doping $p$ decreases through $p^*$\cite{BADOUX}. This is perhaps the most direct evidence so far that the pseudogap opens at a critical point in the normal state. The drop in Hall number can be explained very well\cite{STOREYHALL} in terms of a reconstruction of the large cylindrical Fermi surface into small elliptical pockets, such as those arising from the model of Yang, Rice and Zhang\cite{YRZ} or folding of the Brillouin zone along the (0,$\pm\pi$) to ($\pm\pi$,0) diagonals from antiferromagnetic order\cite{CHUBUKOV}.

Recently, a systematic upturn in the field-induced normal-state resistivity $\rho$ of La-based cuprates has been observed below $p^*$\cite{DAOU2008,LALIBERTE2016,COLLIGNON}. When plotted as the quantity $n_\rho=(1+p)\rho_0/\rho(0)$, where $\rho(0)$ is the zero-temperature intercept and $\rho_0$ is the extrapolated intercept from the high-temperature linear region, the upturn is quantitatively consistent with the same drop in carrier density from $1+p$ to $p$\cite{LALIBERTE2016,COLLIGNON}. Here we will analyze this latest finding using the $\textbf{Q}=(\pi,\pi)$ antiferromagnetic-order-driven Fermi surface reconstruction model\cite{CHUBUKOV} that successfully describes the Hall number\cite{STOREYHALL}. However, it is reasonable to expect similar results from other models possessing similar Fermi surface pockets, e.g. \cite{CHAKRAVARTY,YRZ,HARRISON,SACHDEV,AMPEREAN}. As before we will neglect the field-stabilized subsidiary charge-density-wave order centered around $p$=0.125\cite{LALIBERTE,BADOUXPRX}, for which a further reconstruction of the pseudogapped Fermi surface has been proposed\cite{HARRISON2012}. The constraints imposed by the resistivity data provide new insights into the doping dependence of the mean free path at zero temperature.

\section{Theory}
We begin with a tight-binding energy-momentum dispersion that reproduces the large cylindrical cylindrical Fermi surface observed in well overdoped cuprates\cite{PLATE,VIGNOLLE} centered at ($\pi,\pi$),
 \begin{equation}
\xi_\textbf{k}=-2t(\cos k_x+\cos k_y)
-4t^\prime\cos k_x\cos k_y
-2t^{\prime\prime}(\cos 2k_x+\cos 2k_y)-\mu(p)
 \label{eq:xi}
 \end{equation}
The tight binding coefficients $t^\prime/t=-0.20$ and $t^{\prime\prime}/t=0.134$.
The branches of the $\textbf{Q}=(\pi,\pi)$ reconstructed dispersion are given by
\begin{equation}
E_\textbf{k}^\pm=\left(\frac{\xi_\textbf{k}+\xi_\textbf{k+Q}}{2}\right)\pm\sqrt{\left(\frac{\xi_\textbf{k}-\xi_\textbf{k+Q}}{2}\right)^2+E_g^2(p)}
\label{eq:EK}
\end{equation}
which are weighted by
\begin{equation}
W_\textbf{k}^\pm=\frac{1}{2}\left[1\pm\frac{(\xi_\textbf{k}-\xi_\textbf{k+Q})/2}{\sqrt{[(\xi_\textbf{k}-\xi_\textbf{k+Q})/2]^2+E_g^2(p)}}\right]
\label{eq:WK}
\end{equation}
Here the doping-dependent pseudogap energy $E_g$ represents the energy scale of the antiferromagnetic order. It is set to $E_g(p)=15.4t(0.2-p)$ for $p\leq 0.2$, and is zero for $p>0.2$\cite{STOREYHALL}. The chemical potential $\mu(p)$ is chosen according to the Luttinger sum rule, with the values used in this work listed in table~\ref{MUTABLE}. The hole doping is defined from the area $A_{FS}$ enclosed by the Fermi surface by $2A_{FS}/\pi^2-1$ for the large Fermi surface, and $(Ah_{FS}-Ae_{FS})/\pi^2$ for the reconstructed hole- ($h$) and electron-like ($e$) pockets. 

Finally, the Boltzmann-type conductivity\cite{KONDOTEP} at zero temperature is given by
\begin{equation}
\sigma_{ab}(0) = \frac{e^2}{V}\sum_{\textbf{k},\alpha=\pm}{W_\textbf{k}^\alpha v_x^\alpha(\textbf{k})\ell(p)\delta(E_\textbf{k}^\alpha-E_F)}
\label{eq:Sigma}
\end{equation}
where $v_x^\pm(\textbf{k})=\partial E_\textbf{k}^\pm/\partial k_x$, $\ell$ is the mean free path, $V$ is the normalization volume and $E_F$ is the Fermi energy.

\begin{table}
\caption{Values of $\mu(p)$ used in this work.}
\label{MUTABLE}
\vspace{10pt}
\centering
\begin{tabular}{cc|cc}
\br
$x$ & $\mu(p)/t$ & $x$ & $\mu(p)/t$\\
\mr
0.10 & -1.447 & 0.18 & -0.920\\
0.12 & -1.257 & 0.19 & -0.951\\
0.14 & -1.087 & 0.20 & -0.974\\
0.16 & -0.949 & 0.22 & -1.018\\
\br
\end{tabular}
\end{table}

\section{Results}
\begin{figure}[H]%
\centering
\includegraphics[width=8cm]{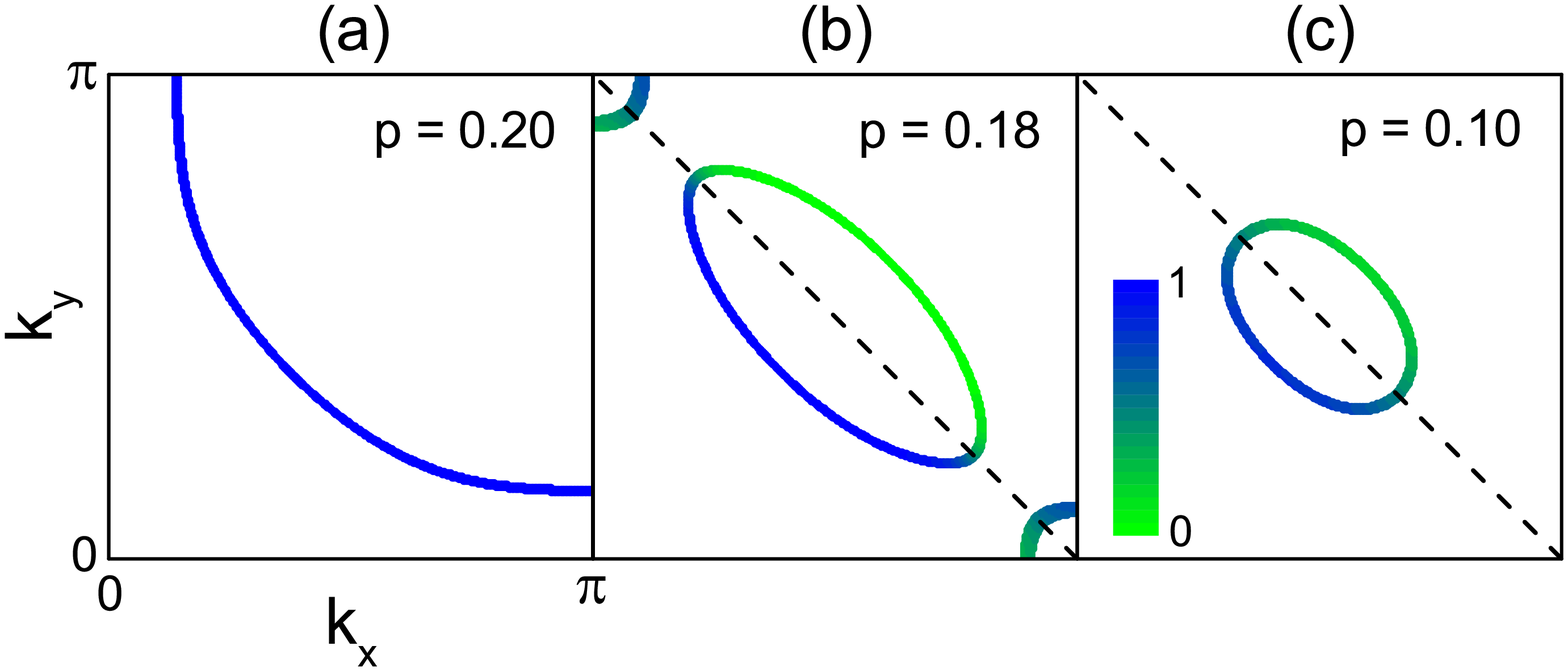}%
\caption{(a) Large tight-binding and (b) \& (c) $\textbf{Q}=(\pi,\pi)$ reconstructed Fermi surfaces for dopings $p$ = 0.20, 0.18 and 0.10 respectively, colored by the weight function $W_\textbf{k}^\pm$ (\ref{eq:WK}).}%
\label{fig:FS}%
\end{figure}

Fermi surfaces for selected dopings calculated from the model above are shown in figure~\ref{fig:FS}. 
In the model, reconstruction of the large hole-like Fermi surface (figure~\ref{fig:FS}(a)) to small nodal hole-like pockets (figure~\ref{fig:FS}(c)) occurs below a critical doping $p^*$=0.20 holes/Cu when our chosen $E_g(p)$ becomes non-zero. In practice, $p^*$ exhibits some material-dependent variability\cite{OURWORK1,BADOUX,LALIBERTE2016,COLLIGNON}. Low spectral weight on the back side of the pockets explains the appearance of arcs in photoemission experiments\cite{KAMINSKI2015}. Note that in a small doping range below $p^*$, $\sim0.17< p < 0.2$, the Fermi surface includes additional small electron-like pockets near ($\pi$,0) and (0,$\pi$) (figure~\ref{fig:FS}(b)) . The extent of this doping range can be tuned by adjusting the rate at which $E_g$ changes with doping\cite{STOREYHALL}. 

\begin{figure}[H]%
\centering
\includegraphics[width=8cm]{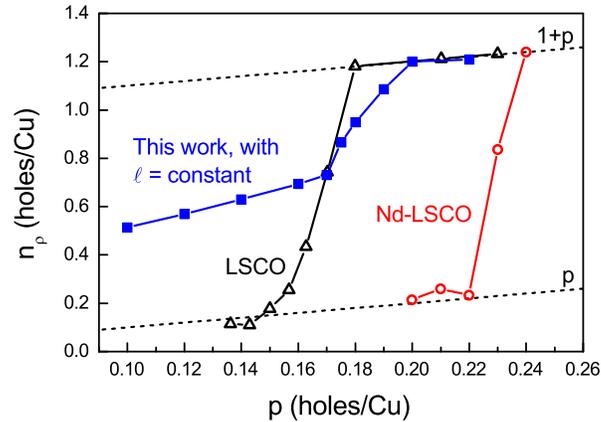}%
\caption{
Calculated conductivity assuming a doping and momentum independent mean free path and rescaled to a carrier density by $n_\rho=1.2\sigma(p)/\sigma(0.2)$ (filled squares)  Experimental values for La$_{1.6-x}$Nd$_{0.4}$Sr$_x$CuO$_4$ (open circles) by Collignon \textit{et al}.\cite{COLLIGNON} and La$_{1-x}$Sr$_x$CuO$_4$ (open triangles) by Lalibert\'e \textit{et al}.\cite{LALIBERTE2016} are shown for comparison.}%
\label{fig:constL}%
\end{figure}

Figure~\ref{fig:constL} shows the conductivity expressed in terms of a carrier density via $n_\rho = 1.2\sigma(p)/\sigma(0.2)$ for a doping and momentum-independent mean free path (as used previously in calculations of the Hall number\cite{STOREYHALL} and thermoelectric power\cite{KONDOTEP,STOREYTEP}). Experimental values from high-field resistivity measurements down to 1.5K on La$_{1.6-x}$Nd$_{0.4}$Sr$_x$CuO$_4$\cite{COLLIGNON} with $p^*$=0.235 holes/Cu, and La$_{1-x}$Sr$_x$CuO$_4$\cite{LALIBERTE2016} with $p^*$=0.18 holes/Cu, are shown alongside for comparison. A drop in the calculated conductivity is evident, due solely to the collapse in size of the Fermi surface, however it is only about half the size required to scale with the actual change in carrier density. 

\begin{figure}[H]%
\centering
\includegraphics[width=8cm]{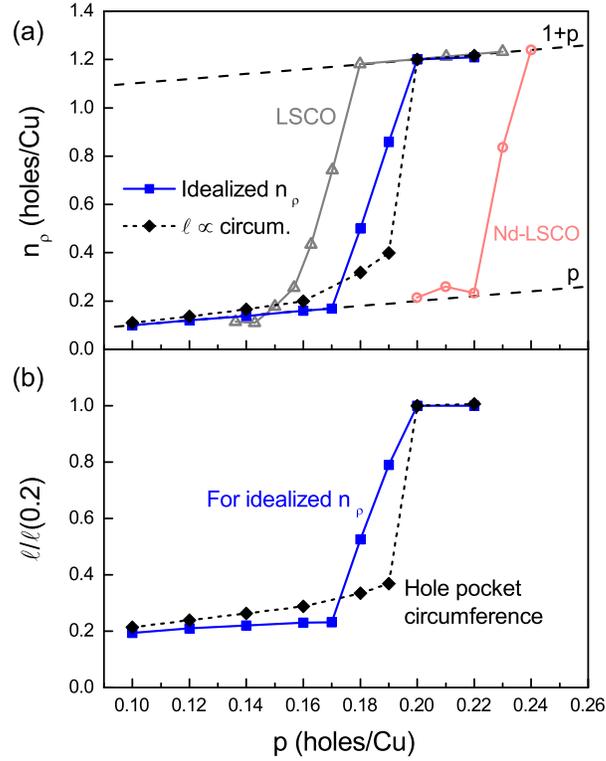}%
\caption{
(a) Idealized conductivity that scales with a drop in carrier density from $1+p$ to $p$ below $p^*=0.20$ (squares). Conductivity assuming a mean free path proportional to the circumference of the hole-like Fermi surface pockets (diamonds). Experimental values from \cite{COLLIGNON,LALIBERTE2016} are shown for comparison (open symbols). (b) Mean free paths that produce the conductivities shown in (a), normalized by their values at $p^* = 0.20$.}
\label{fig:MFP}%
\end{figure}

In order to obtain better agreement with experiment we can take the mean free path as a fitting parameter. In figure~\ref{fig:MFP}(a) we plot an idealized $n_\rho$ that transitions from $1+p$ to $p$ below $p^*$. The requisite mean free path is shown below in figure~\ref{fig:MFP}(b). The figure demonstrates that in order to achieve a change in conductivity that scales with the carrier density the mean free path must collapse below $p^*$. 

\section{Discussion and Conclusions}

The calculations suggest a close link between the mean free path and the size of the Fermi surface. Meanwhile the experimentally observed relation between the resistivity upturn and carrier density implies that the conductivity is proportional to the area enclosed by the Fermi surface. At zero temperature, the conductivity can be expressed as an integral of the mean free path around the Fermi surface\cite{KONDOTAU}. For a circular 2D free-electron-like Fermi surface, integrating a mean free path proportional to the circumference $2\pi k_F$ yields a conductivity proportional to $k_F^2$ and hence the area enclosed. Inspired by this, we calculate the conductivity (figure~\ref{fig:MFP}(a)) using a mean free path proportional to the circumference of the hole-like Fermi surface pockets (figure~\ref{fig:MFP}(b)) and achieve very reasonable agreement with the idealized case. 

The circumference itself is proportional to the average chord of the pocket. This points to intrapocket scattering as a possible driver of the upturn in resistivity, facilitated by out-of-plane defects which act primarily as small-momentum transfer (or forward) scatterers\cite{SCALAPINO2006,GRASER,ALLOUL2009}. The increase in forward scattering due to the small pocket size apparently overcomes any decrease in the density of scattering final states due to truncation of the Fermi surface\cite{OVERHAUSER}, that would otherwise lead to a decrease in resistivity.

Compared to the idealized case, a more abrupt transition from $1+p$ to $p$ arises because the antinodal electron-like pockets were neglected. Those pockets account for the roughly linear transition in the Hall number from $1+p$ to $p$ between $p$=0.2 and 0.165\cite{STOREYHALL}. It is likely that they would play a similar role in $n_\rho$, though their associated mean free paths would need to be quite long. Interpocket scattering, perhaps facilitated by the same $\textbf{Q}=(\pi,\pi)$ wave vector behind the Fermi surface reconstruction, would be worth exploring. A more detailed treatment of the transition region is also of relevance to ongoing analysis of the temperature dependence of the resistivity and the question of whether the pseudogap closes at $T^*$\cite{OURWORK1,STOREY2015}.

How does this doping dependent mean free path affect the results of previous Hall effect\cite{STOREYHALL} and thermopower\cite{STOREYTEP} studies which assumed it is constant? Since both of these quantities depend on a ratio of conductivities\cite{HOPFENGARTNER,KONDOTEP}, the magnitude of the mean free path cancels out and those results effectively remain unchanged. 

During peer review of this article, work by Chatterjee \textit{et al.}\cite{CHATTERJEE2017} was brought to the author's attention which reports detailed calculations of the thermal and electrical conductivities near quantum phase transitions where static or fluctuating antiferromagnetic order vanishes. Their work considers the scattering rate $\tau^{-1}$ rather than the mean free path $\ell$, the two being related through the velocity by $\ell=v\tau$. The conclusion is similar. A constant scattering rate does not produce the observed change in conductivity. Instead an increase in scattering rate is required below $p^*$, which is proposed to arise from patches of density waves.

\ack
Supported by the Marsden Fund Council from Government funding, administered by the Royal Society of New Zealand. The author acknowledges helpful discussion with J.L. Tallon.

\end{document}